\documentclass[revtex4-1, nofootinbib, 10pt, twocolumn, floatfix, groupedaddress, aps, longbibliography] {revtex4-1}

\usepackage{times, mathrsfs, amsmath, amsfonts, graphics, graphicx, cancel, color, amsthm, bbm, mathtools, amssymb, physics}
\usepackage[nice]{nicefrac}
\usepackage[english]{babel}
\usepackage[margin=.7in]{geometry}
\usepackage[T1]{fontenc} 
\usepackage{capt-of}

\usepackage{xfrac,relsize,float}
\usepackage{comment}
\usepackage{slashbox}
\newcommand{\cmark}{\ding{51}}%
\newcommand{\xmark}{\ding{55}}%
\usepackage{appendix}
\usepackage{amssymb}
\usepackage{pifont}
\usepackage[unicode=true, breaklinks=false, pdfborder={0 0 1}, backref=false, colorlinks=true, linkcolor=blue, citecolor=blue]{hyperref}

\def\ket#1{\mathinner{|{#1}\rangle}}

\def\BraVert{\egroup\,\mid\,\bgroup}

\definecolor{Blue}{rgb}{0,0,1}
\definecolor{Red}{rgb}{1,0,0}
\definecolor{Green}{rgb}{0,1,0}
\definecolor{darkgreen}{rgb}{0,.7,0}
\definecolor{Purp}{rgb}{.2,0,.2}
\definecolor{white}{rgb}{1,1,1}

\usepackage{bbold}

\begin{document}
\title{The relativity of indeterminacy}

\author{Flavio Del Santo}
\affiliation{
Institute for Quantum Optics and Quantum Information (IQOQI-Vienna), A-1090 Vienna, Austria; and
Faculty of Physics, University of Vienna, A-1090 Vienna, Austria}
\author{Nicolas Gisin}
\affiliation{Group of Applied Physics, University of Geneva, 1211 Geneva 4, Switzerland; and SIT, Geneva, Switzerland}

\date{\today}

\begin{abstract}
A long-standing tradition, largely present in both the physical and the philosophical literature, regards the advent of (special) relativity --with its block-universe picture-- as the failure of any indeterministic program in physics. On the contrary, in this paper, we note that upholding reasonable principles of finiteness of information hints at a picture of the physical world that should be both relativistic and indeterministic. We thus rebut the block-universe picture by assuming that fundamental indeterminacy itself should  as well be regarded as a relational property when considered in a relativistic scenario. We discuss the consequence that this view may have when correlated randomness is introduced, both in the classical case and in the quantum one.
\end{abstract}
\maketitle

\section{Relativity from information principles}

In Newtonian mechanics the interaction of particles is described by the potential energy of interaction which is a function of only the positions of the interacting particles, hence, as stressed also by Landau and Lifshitz,  such a theory ``contains the assumption 
of instantaneous propagation of interactions'' \cite{landau}. They thus motivate the development of the theory of special relativity (SR) as to overcome a long-standing problem of ``infinities'' (the infinite speed of propagation of the interactions in this case), which affected classical physics.\footnote{Note that SR only deals with mechanics and electromagnetism, whereas to address similar problems about gravity one has to consider general relativity.}

In recent years, new attention has been brought to the problem of having infinite quantities in physical theories \cite{ellisinf, lynds}, with a special emphasis on the connection between physics and information \cite{svozil, gisin1, delsantogisin, dowek, drossel, NGHiddenReals, ben, delsanto, gisin2020}.\footnote{Many of these approaches stem from the assumption of Landauer's principle according to which, in short, ``information is physical'' \cite{landauer}.} In particular, in Refs. \cite{gisin1, delsantogisin, NGHiddenReals, gisin2, delsanto, gisin2020}, we have pointed out that the usual understanding of classical (non-relativistic) mechanics rests on the unwarranted implicit assumption of  ``infinite precision'', namely, that every physical quantity has an actual value with its \emph{infinite} determined digits (which formally translates into the assumption that physical variables take values in the real numbers). To overcome this problem of ``infinities'', we have assumed the ``finiteness of information density'' (i.e., finite regions of space cannot contain but finite information) as a foundational principle, and showed how this leads to fundamental indeterminism even in classical physics.

Remarkably, also the problem of infinities raised by Landau and Lifshitz can be addressed by invoking the same principle of ``finiteness of information density'', thus assuming it as an alternative axiom from which deriving SR. Consider the following two axioms:
\begin{description}
\item[P1 -- Principle of relativity] The laws of physics have the same form in every inertial frame of reference.
\item[P2 -- Principle of finiteness of information density] A finite volume of space can only contain a finite amount of information.
\end{description}

From P2 one infers that only a finite amount of information can be transmitted in a finite time, otherwise this would require to ``move'' an infinite volume of space. Hence, the \emph{signal velocity} (i.e. the speed of propagation of information) necessarily needs to be finite too. Now, if there is a maximal velocity,\footnote{In principle a finite velocity could be unbounded, but we restrict our analysis  to the case where there exists a maximal velocity.} from P1, this must be the same in every inertial reference frame \cite{landau}. The former considerations are enough to derive the whole theory of special relativity (possibly with the further but quite innocuous assumptions of homogeneity of space and time and isotropy of space). Incidentally, note that the fact that this maximal signal velocity is identified with the speed of light in vacuum does not follow from physical principles, nor is it required at all to derive SR.\footnote{However, a body moving at the maximal signal velocity would be required by SR to be massless (see e.g. \cite{morin}, p. 589). This, together with the empirical demonstration that Maxwell's equations are invariant under Lorentz transformations, leads to the identification of the maximal velocity with the speed of light in vacuum, $c$. Note that $c$ is not only identified with the speed of light but it also appears as a natural constant in several other theoretical frameworks (see \cite{ellislight}), most notably as the speed of propagation of gravitational waves in vacuum. The numerical equivalence of these speeds has been recently empirically verified with high accuracy \cite{grav}.}

The Principle of finiteness of information density (P2) has thus two physical consequences: On the one hand, it imposes that, in general, physical quantities (say the position of a particle at a certain moment) do not have perfectly determined values at every time (or, alternatively, that the truth value of certain empirical statements, such as ``a particle is located at $\bar{x}$  at a certain instant $\bar t$'', is indeterminate). On the other hand, P2 also imposes a dynamical limit to the propagation of information (signal velocity) which, together with P1, allows to fully derive the theory of SR. Hence, \textbf{upholding the Principle of finiteness of information density gives us a hint that physics should be at the same time indeterministic and relativistic}. But are these two worldviews compatible? 
To answer this question, in what follows we rephrase, by means of a physically intuitive scenario, a by now classical argument independently put forward by Rietdijk \cite{Rietdijk66} and Putnam \cite{Putnam67}, that ``illustrate[s] the power of the block-universe picture'' \cite{smolin} by allegedly showing the incompatibility between (special) relativity and indeterminism. We then discuss its implicit assumptions, pointing out that one of them --which seems innocuous in Newtonian physics--  is problematic in special relativity, thus making their argument untenable. As we shall see, the consequences of putting together the concepts of indeterminism and relativity would lead us to conclude that the state of determinacy of a physical (random) variable is relative as well. The relativity of indeterminacy will be analyzed in the different scenarios of classical independent randomness, classically correlated randomness and quantum correlated randomness.

\section{Indeterminacy is relative}
\label{rel indet}
\subsection{Locally and independently generated randomness}
\label{class rand}
To discuss indeterminism in the framework of SR, and rephrase Rietdijk's and Putnam's argument, we introduce the concept of a True Random Number Generator (TRNG), by which we mean an abstract device that outputs genuinely  random bits. Namely, before each bit is output, its value is not only unknown (epistemic uncertainty), but it actually has no determined value (ontic indeterminacy), even though there might be a probability distribution associated to the outcome to be realized (in principle, both at the epistemic and at the ontic level). More precisely, even having complete knowledge of (i) the \emph{state}, i.e. the values of all the variables that may influence the outcome of the TRNG (this can be in principle everything that lies in the past light cone of the event associated to the generation of the bit), and (ii) of the \emph{dynamical laws} that rule the evolution of each and every of said variables, there is no way to predict with certainty which will be value of the bit output by the TRNG, not even in principle (see Table I). 

\vspace{0.5cm}
\begin{tabular}{|*{5}{c|}}\hline
\backslashbox{Ontic}{Epistemic}
&Known $a$ & Unknown $a$\\\hline
Determinate $a$ & \cmark& \cmark \\\hline
Indeterminate $a$ &\xmark & \cmark\\\hline
\end{tabular}
\captionof{table}{\small{Differences between epistemic (un)certainty and ontic (in)determinacy}. The value of a variable $a$ can be known only if it is determinate.}
\vspace{0.5cm}

This is what we mean by true randomness, which clearly entails indeterminism. In what follows, we will be concerned solely with the (ontic) indeterminacy that each bit has before it is output and acquires a definite value (on the contrary, we will not consider what dynamical laws --deterministic or otherwise-- govern the evolution of the considered systems). This very concept can be expressed in terms of truth values of empirical propositions, such that the statement, say, ``the value of the  bit $j$ output by the TRNG at time $\bar t$ (in a certain reference frame) is $j=0$'' has a definite truth value, either true or false, only after $\bar t$, whereas it was (ontologically) indeterminate before, i.e. its truth value is neither true nor false.\footnote{We refer to propositions (or statement) as \emph{empirical} because they are about the values taken by physical variables, i.e. outcomes of hypothetical experiments. However, in principle, we will not immediately disregard also those statements that refer to not directly empirically accessible quantities (e.g., a joint proposition about two outcomes of experiments conducted in space-like separation), which instead would be meaningless to genuine empiricists  because they are unverifiable.} Note that this makes the law of the excluded middle of classical logic fail, and it so relates the concept of indeterminacy to mathematical intuitionism (see  \cite{gisin2020, posy}).

\begin{figure}[h!]
\centering
\includegraphics[width=8.7cm]{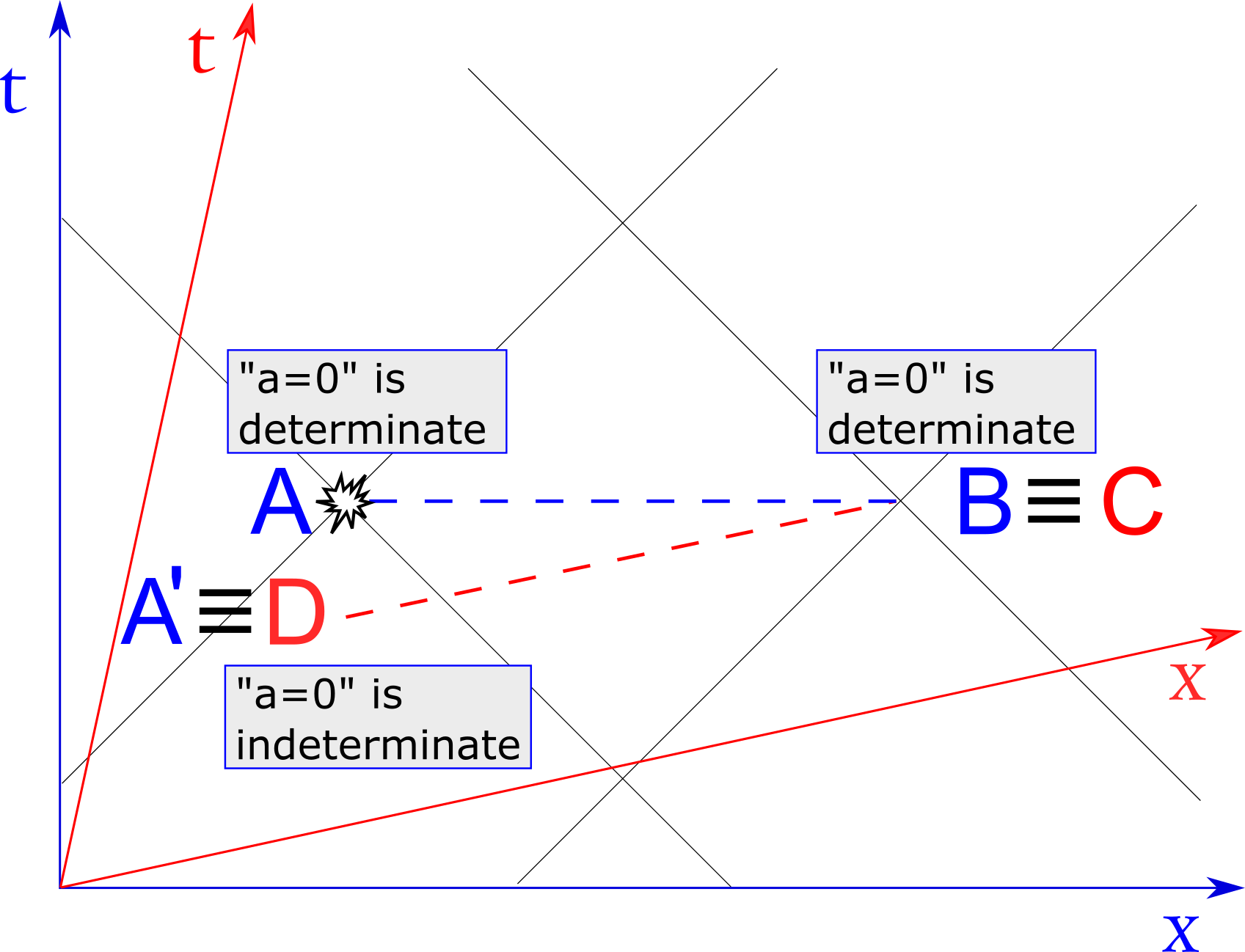}
\caption{\small{Space-time diagram (in 1+1 dimensions) illustrating Rietdijk's and Putnam's argument for the alleged incompatibility between special relativity and indeterminism. 
The observers Alice (A) and Bob (B) are at rest in the blue reference frame, whereas Charlie (C) and Debbie (D), also at relative rest, move at a constant speed in the positive $x$ direction (their transformed reference frame is depicted in red). The blue dotted line represents the plane of simultaneity in Alice's and Bob's rest frame at the instant in which Alice's True Random Number Generator outputs the bit $a$ (i.e. when it becomes determinate; event A). In the moving reference frame (red), however, when Charlie overlaps with Bob, he is simultaneous with Debbie, which in turn overlaps with Alice's past (event A') when $a$ was not yet determinate.}}
\label{fig1}
\end{figure}
Turning now to SR, with reference to Fig. \ref{fig1}, let us consider four inertial observers, Alice, Bob, Charlie and Debbie. Assume that Alice has a TRNG that outputs a fresh random bit every minute (locally in her inertial reference frame). Assume also that Bob, at rest in Alice's frame, is located at a certain fix distance from Alice, say, one minute away at light speed (one light-minute). Charlie and Debbie, at relative rest, move at a constant speed $v$ with respect to Alice's and Bob's frame. All parties know that Alice's TRNG outputs a fresh random bit every minute in her frame, i.e. every $1/\gamma=1/\sqrt{1-v^2/c^2}$ minutes in Charlie's and Debbie's reference frame. Moreover, the positions and velocity $v$ are tailored such that at 1:00 pm in both reference frames Charlie's position lies on Bob's world-line (point $B\equiv C$ in Fig. \ref{fig1}), while Debbie's position overlaps with Alice's world-line (point $A' \equiv D$ in Fig. \ref{fig1}).

Let us denote by $a\in \{0,1 \}$ the bit output by Alice's TRNG at 1:00 pm in her reference frame. As explained above, by the very nature of a TRNG, at any time instant before 1:00 pm, the proposition ``$a=0$'' has no truth value for Alice (nor for any possible observer for that matter) but rather is fundamentally indeterminate. Yet, at 1:00 pm, Alice's TRNG outputs a fresh bit that now gets determined for Alice. Is the value of this bit determined for Bob, too? One may be tempted to answer in the positive (as in fact Rietdijk and Putnam do), since the events happening at Alice's and Bob's locations, respectively, are simultaneous in their rest frames. Of course, Bob may not know the value of $a$ because Alice may not have communicated it, or, more fundamentally, because Bob is space-like separated from Alice. But does the proposition ``$a=0$'' have a truth value or is it indeterminate for Bob? 

Since at 1:00 pm Bob and Charlie are next to each other --ideally they exactly overlap at the same location-- it is plausible to assume that the proposition ``$a=0$'' has the same truth value for both of them (since this is empirically testable there is actually no room for any alternatives at the ontological level). Hence, let's assume it has a determinate truth value for Charlie as well. But then, by applying the same reasoning as before, this implies that the proposition also has a truth value at all locations simultaneous to Charlie in his inertial frame, thus it must be determinate for Debbie too. Yet, given the relative motion between the two reference frames, Debbie overlaps with Alice's world line at the space-time point $A'$ where her TRNG has not yet output the bit. Hence, following this chain of inferences,  ``$a=0$'' had already a definite truth value before Alice's TRNG outputs the bit $a$, which is a contradiction with the assumption of a TRNG. From this contradiction, Rietdijk and Putnam concluded that, in general, the structure of SR does not allow for indeterminate events. 

The previous argument, however, relies on two assumptions that seem \emph{prima facie} innocuous, which originate from an intuitive extension of classical concepts to a relativistic scenario. In fact, these assumptions are introduced to characterize what it means for two (or more) observers to share a determinate reality (in terms of truth values of propositions). Their assumptions can be made explicit as follows:
\begin{enumerate}
\item \emph{Local reality}: Any two observers that locally overlap attribute the same truth values to empirical propositions (including the value ``indeterminate''). 
\item \emph{Present reality}: Any two distant observers at relative rest attribute the same truth values (including the value ``indeterminate'') to empirical propositions about present events (i.e., lying on the same plane of simultaneity in their rest frame).
\end{enumerate}
However, while the former of these assumptions may still be upheld in SR because any two observers can operationally verify the consistency of their attributed truth values locally, the latter assumption becomes questionable. Indeed, we maintain that the finiteness of the maximum speed of propagation of any piece of information renders also determinacy (i.e. the definiteness of the truth values of empirical statements) a relational property. This can be seen by taking any binary function (e.g. the sum modulo 2) of the statements describing the outcomes of local physical processes taking place at distant locations (e.g. the output bits of two distant TRNG's). It thus becomes a straightforward assumption that determinacy itself propagates in space at the maximal signal velocity, i.e. instantaneously in Newtonian physics, but at a finite speed (that of light) according to relativity. This is where the Rietdijk's and Putnam's argument fails, in the assumption that truth values are always shared by observers lying on the same plane of simultaneity, even  though this is not verifiable except for the region enclosed within the intersection of their future light cones. Hence, in short, (in)determinacy is relative.
\begin{figure}[h]
\includegraphics[width=8.7cm]{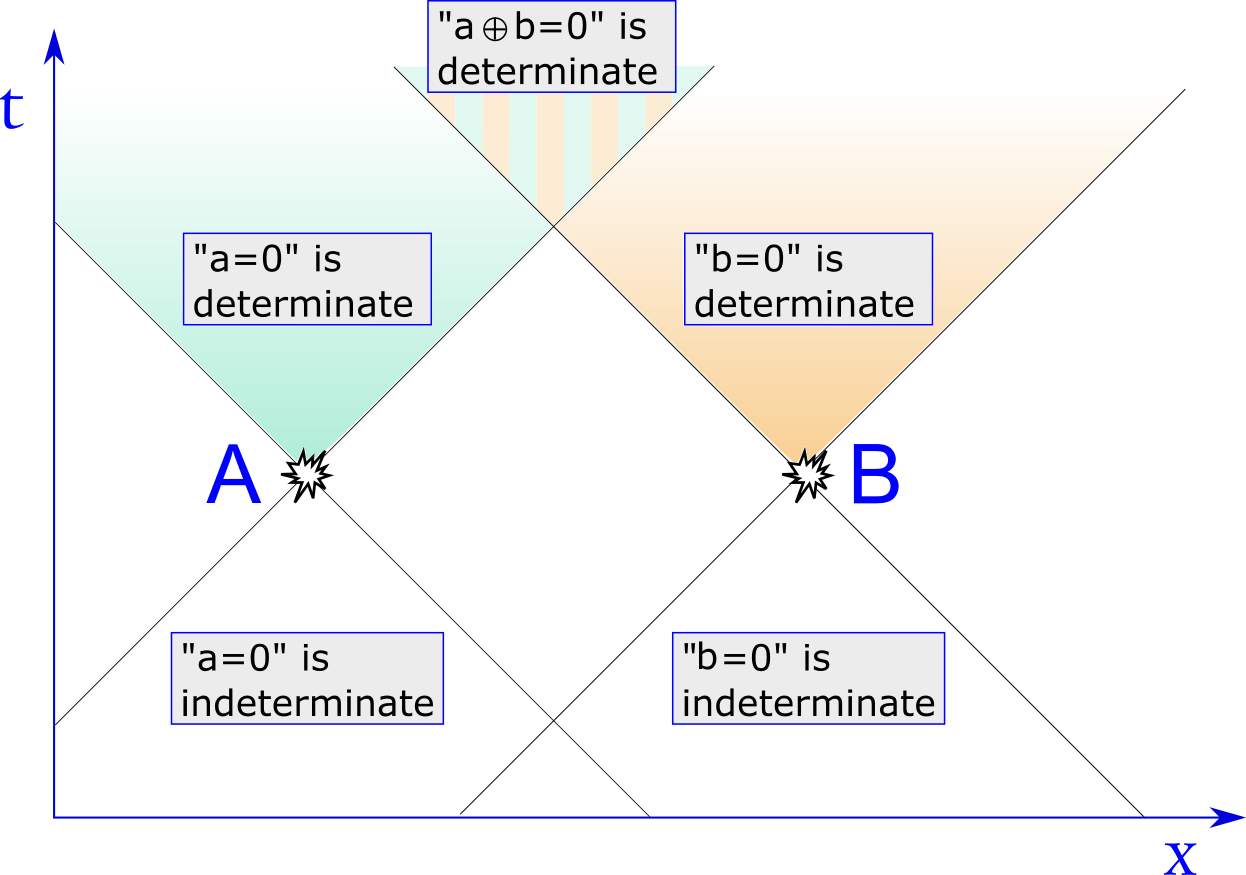}
\caption{\small{Space-time diagram (in 1+1 dimensions) showing that for distant observers (in)determinacy is relative. Even if each of their local TRNG outputs a bit, becoming determinate from indeterminate, it is only in the overlap of their future light cones that both bits become determinate. Note that both $a=0$ and $b=0$ are indeterminate in the entire white region.}}
\label{fig2}
\end{figure}

In our simple example, this means that at 1:00 pm the proposition ``$a=0$'' has no definite truth value for Bob, nor for Charlie, but it is for them still  indeterminate, as it was for Alice before 1:00 pm. It is only one minute later, i.e. at 1:01 pm, that ``$a=0$'' acquires a definite truth value for Bob, though Bob may still not know this value. For instance, if Bob also holds a TRNG --here assumed to generate randomness locally and independently of Alice's TRNG-- which also outputs a fresh random bit, denoted by $b$, every minute, then the value of their sum modulo 2, $a\oplus b$, is indeterminate everywhere until 30 seconds after 1:00 pm (in their common inertial frame), and the proposition ``$a=b$'' has no truth value, i.e. is indeterminate, until then (see Fig. \ref{fig2}).

Note that the present rebuttal of Rietdijk's and Putnam's argument is in agreement with those put forward by Stein and Savitt, respectively \cite{Stein91,Savitt09}.

\subsection{``Classical'' correlated randomness}
In the previous section, we have argued that contrarily to a well-known philosophical argument, indeterminism and relativity remain compatible, insofar as the (in)determinacy of the values taken by physical variables, each measured by a distinct relativistic observer, is relative. So far, however, we limited our analysis to the assumption that each TRNG generates random bits at its local position and independently from any other random generators. Yet, one can in principle envision a form of ``classical'' correlated randomness between two or more TRNG's, which will help bridge the gap between the classical and the quantum case.\footnote{Here by ``classical'' we mean without introducing the structure of quantum probabilities, but merely consider systems whose outcomes are correlated through randomness.}

Let us thus consider once more two distant observers, Alice and Bob, each provided with a TRNG such that their respective bits $a$ and $b$ take one of their two possible values with equal probability, i.e. $p(a=0)=p(a=1)=\frac{1}{2}=p(b=0)=p(b=1)$. However, when considered jointly, some correlations between the bits output by the two TRNG may have been established.\footnote{We do not discuss here possible physical mechanisms that may establish this  kind of correlations, but we assume that in principle these can exist.} For example, although at the local level the values of the bits are random with uniform distribution, there could be a bias towards certain joint results. For instance, the cases when their values are the same could occur with a higher probability, say $p(a=b)=\frac{3}{4}$.  One can think that at the ontological level there is a physical property such as a (non-local) propensity --i.e. an objective tendency that quantifies the bias towards the possible realization of an indeterministic outcome-- that accounts for the correlations between the two TRNG.\footnote{This is inspired by the propensity interpretation of probability which was introduced by Popper \cite{popper57}; see also \cite{delsantogisin} for a recent application of this concept to indeterministic physics.}
Due to the space-like separation between Alice and Bob, there exist inertial reference frames in which Alice's TRNG generates a random bit, say $a=0$, before Bob's one (which thus remains indeterminate in such reference frames). But because of the correlations between TRNG's, within the future light cone of Alice, the propensity for Bob to find the outcome 0 gets updated from $p(b=0)=\frac{1}{2}$ to $p(b=0|a=0)=\frac{3}{4}$. On the other hand, in a reference frame in which the realization of a fresh bit by Bob's TRNG comes first, it is the propensity for the outcome of Alice's TRNG that gets updated according to the established correlations. Note that the realization of a fresh bit by a local TRNG, say, on Alice's side, does not directly affect the value of the bits generated on Bob's side (otherwise this could be used to signal). It is Alice's propensity for the outcome of Bob's TRNG that gets updated, such that if she obtains $a=0$, within her future light cone the propensity associated to the statement $b=0$ will change (to 3/4 in this case). In this way, when Alice's bit gets determinate, also the propensity for the  value of $b$ relative to Alice's future light cone gets determinate, but not the value of $b$.  It is only within the intersection of the future light cones of Alice and Bob that their outcomes unambiguously assume a definite value which ought to comply with this non-local propensity.

The reader acquainted with quantum physics would have already noticed the similarity of this ``classical'' non-local randomness with quantum correlations. However, we wanted here to express a hypothetical property of indeterminacy which can be conceptually introduced independently of quantum theory (although we do observe it experimentally in quantum systems and not in classical ones). Before moving to the quantum case, we deem it useful to consider the so-called Popescu-Rohrlich (PR) boxes \cite{pr}, because, contrarily to the indeterministic scenarios so far discussed (in terms of TRNG), they admit the choice of inputs that is crucial in the quantum case (i.e., the choice of the measurement basis). PR boxes are a theoretical model of an operational setup that displays the maximal amount of correlated (non-local) randomness which however does not lead to instantaneous signaling (i.e. it respects the no-signaling conditions). In terms of correlations between two distant parties --which carry out local operations (i.e. they chose an input bit $x$ and $y$, respectively) and measure an outcome of some not fully specified random process (indicated by the bits $a$ and $b$, respectively)-- the no-signaling conditions are given by the fact that the input choice (which in turn could be picked according to local and independent TRNGs) of one party cannot directly influence the outcome of the other one. PR boxes are a set of correlations $p(a,b|x,y)$ defined as follows: if the pair of inputs, $(x,y)\in \{ (0,0),(0,1),(1,0)\}$, then the outputs are perfectly correlated, i.e., $p(0,0|x,y)=p(1,1|x,y)=1/2$. Otherwise, if the input pair $(x,y)=(1,1)$, the outcomes are perfectly anti-correlated, i.e.,  $p(0,1|1,1)=p(1,0|1,1)=1/2$. It is straightforward to show that these correlations respect the no-signaling conditions. Accordingly, if, for example, $x=0$ and $a=0$, then the propensity $p^A(b|x,y,a)=1$ (i.e., in this case  $b=0$), where the superscript index indicates Alice's future light cone, but $p(b|y)=1/2$ everywhere else.

\subsection{Quantum correlated randomness}
In the previous sections, we have discussed different ``classical'' scenarios that bring together true randomness (i.e., indeterminism) and special relativity, reaching the conclusion that (in)determinacy is relative. Moreover, in the case of correlated randomness between two TRNGs, there are regions of space-time in which different inertial observers attribute, in general, different  probabilities for measurement outcomes (corresponding to different objective propensities), and it is only in the overlap between their future light cones that their predictions match (where they also become testable). However, it would be impossible to discuss indeterminism without addressing quantum physics, which not only is the most successful theory ever in terms of predictions, but also a theory that hints at the indeterministic nature of our world. Indeed, the violation of Bell's inequalities \cite{bell} has proven that if there is at least a random event in the universe, then there can be arbitrarily many of them \cite{acin} (see also \cite{delsantogisin} for a discussion), and even that classical trajectories of particles cannot exist predetermined (if one upholds locality) \cite{trajectories}. Moreover, in a recent work, Dragan and Ekert showed that elementary special relativistic considerations can lead to quantum randomness \cite{dragan}.

Hence, we now introduce quantum (non-local) correlations, that is, what happens if Alice's and Bob's TRNGs are entangled? Assume an initial maximally entangled state, say the singlet $\ket{\Psi^-}=1/\sqrt 2 (\ket0 \ket1-\ket1 \ket0)$, where $\ket 0$ and $\ket 1$ are eigenstates of $\sigma_z$, and the first qbit is with Alice and the second with Bob. When Alice measures her local qbit in an arbitrary basis, say the $x$-basis (recall that $\ket{\Psi^-}$ has the same form in every basis), she obtains, for example, the outcome $a=0$, corresponding to the state $\ket 0_x$. While Bob's local quantum state remains unchanged,  Alice describes, after her measurement, the global state using the projection postulate, i.e. it becomes $\ket0_x\ket1_x$. Symmetrically, however, Bob locally measures his qbit, which is still in the initial entangled state $\ket{\Psi^-}$, in the $y$-basis, and obtains, say, outcome $b=0$, corresponding to $\ket 0_y$. By using the projection postulate, he updates the global state to $\ket1_y \ket0_y$, while Alice's state remains unchanged. Since Alice's and Bob's measurements are carried out in space-like separation, there are inertial reference frames in which one measurement occurs before the other and vice versa, from which we conclude that there are certain regions of space-time in which the global quantum state as assigned by Alice and Bob differ. At the intersection of the respective future light cones, however, the two bits acquire their value and the quantum state is reduced to $\ket0_x\ket0_y$, which reconciles with the usual projection postulate. In the way here described, the state is well defined at every point in space-time, but in every inertial frame there are regions of space-time where two different states are attributed to the two qbits. Notice that the probabilities of the two outcomes $a$ and $b$ are correlated, as usual in quantum mechanics (if one performs local measurement in two bases which are not mutually unbiased): this is non-local randomness \cite{gisinbook}, but the states change only locally. 

\begin{figure}[h!]
\includegraphics[width=9cm]{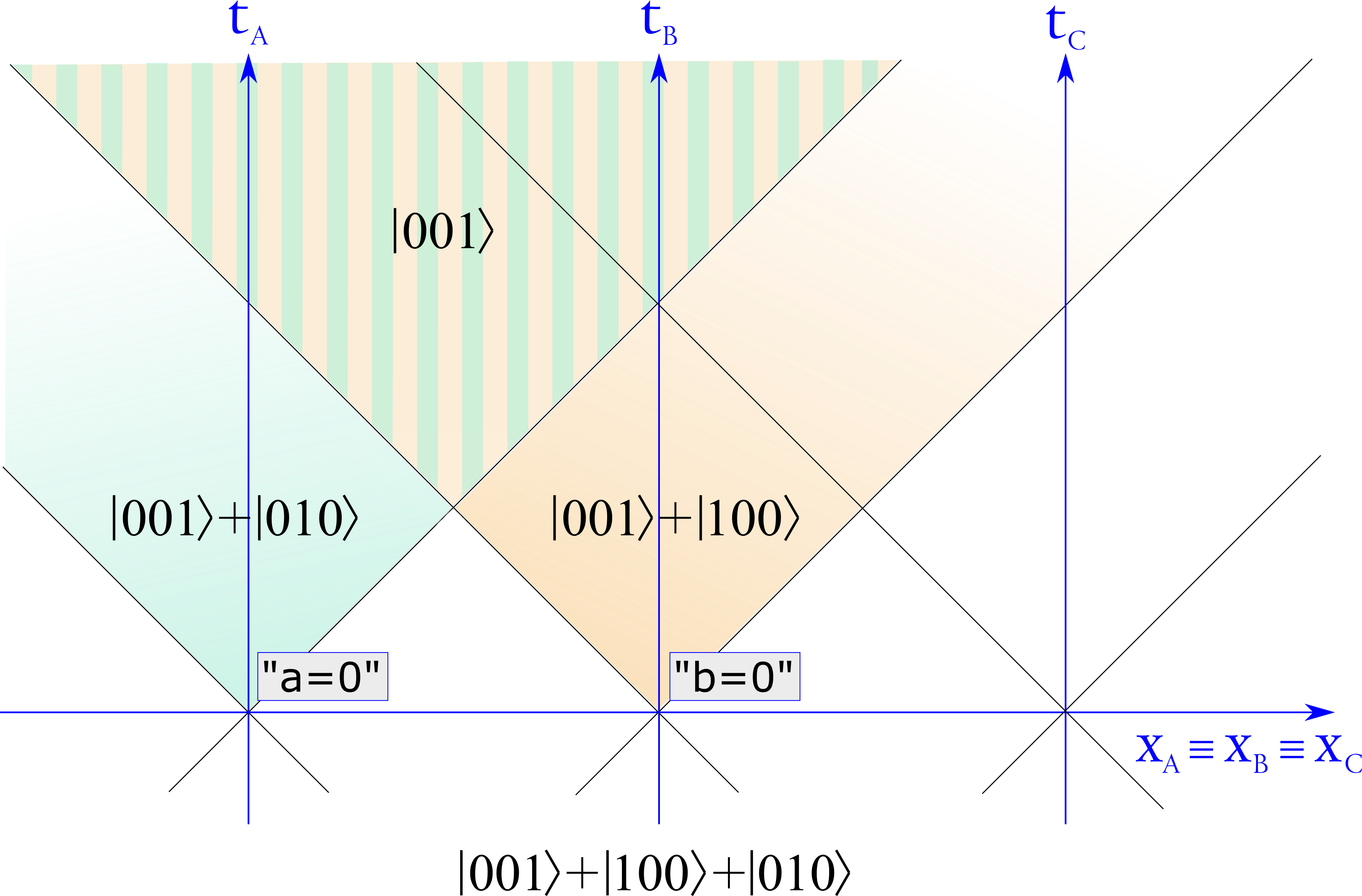}
\caption{\small{Space-time diagram (in 1+1 dimensions) showing a relativistic scenario (readapted from Ref. \cite{aharonov84}) in which different global quantum state are assigned in different regions of space-time.}}
\label{fig3}
\end{figure}
Our view is similar to that of Aharonov and Albert, who, in a series of papers \cite{aharonov81, aharonov84, aharonov86}, pointed out the inadequacy of the standard quantum state-vector description for multipartite systems in a relativistic scenario. We thus deem worth to rephrase, in the language of the present paper, one of the examples they gave in Ref. \cite{aharonov84}. With reference to Fig. \ref{fig3}, let Alice, Bob and Charlie be located at three distant locations and share a tripartite entangled W-state $\ket{\alpha}\propto \ket1\ket0 \ket0 +\ket0\ket1 \ket0+\ket0 \ket0 \ket1$, (normalization is omitted) which can be seen as the Fock representation of a single particle in a quantum superposition between the three different locations (entanglement with vacuum). In a certain reference frame, a measurement is performed on the first qbit at $t_1$ revealing that the particle is not located at Alice's position (formally this means to apply to Alice's qbit the projector $| 0\rangle  \langle0| $). The global state thus gets updated to $\ket \beta \propto  \ket0(\ket1 \ket0+\ket0 \ket1)$. A second measurement is then performed, at time $t_2$, on the second qbit, revealing that the particles is also not located at Bob's location. This leaves the global state in $\ket \gamma = \ket0\ket0 \ket1$. Given the space-like separation of these events, however, there exists another reference frame, in which the measurements happen in reversed order. In that frame, the initial state is the Lorentz transformed of $\ket{\alpha}$, which we indicate by $\ket{\alpha'}$. The first measurement happens at $t_2'$ and leads to the updated state $\ket{\eta}\propto  \ket{1'}\ket{0'} \ket{0'}+\ket{0'}\ket{0'} \ket{1'})$ and then, after the second measurement at time $t_1'$, to the final global state $\ket{\gamma'} = \ket{0'}\ket{0'} \ket{1'}$. This shows that the intermediate states $\ket \beta$ and $\ket \eta$, which describe the tripartite system in the period in between the two measurements in their respective reference frames, differ in a way that they are not the Lorenz transformed of one another.

\section{Outlook}
\label{fiq rel}
We have argued that, under the reasonable assumption of finiteness of information density, physics should comply with both indeterminism and relativity. Notwithstanding some historical criticisms, we have shown that these two views are compatible if one regards (in)determinacy itself as relative. Our analysis was based on the formulation of indeterminacy in terms of a third logical truth value (besides ``true'' and ``false'') for empirical propositions. This is thus reminiscent of mathematical intuitionism, where the law of the excluded middle fails \cite{gisin2020, posy}. 

Despite these promising preliminary  conclusions, a number of fundamental questions concerning a hypothetical physical framework that would bring together special relativity and indeterminism (possibly independently of quantum mechanics) remain open. For example, if we enforce the principle of finiteness of information density, the concept of a relativistic ``event'', usually defined as a (mathematical) point in the Minkowski space-time, would need to be substituted by a finite hypervolume. This implies that also light cones would not be perfectly determined, but their edges would be somehow blurred; does this mean that the future determinacy of physical variables is also not perfectly determined? To this end, following the principle of finiteness of information density, we have introduced in previous works \cite{gisin1, delsantogisin} a concrete model of indeterministic classical (non-relativistic) physics. This was achieved by replacing the usual real numbers --which are customarily assumed to be the values taken by physical quantities-- with what we named ``finite information quantities'' (FIQs). It would be interesting to analyze in details what would be the consequences of introducing FIQs in the context of special relativity.

Furthermore, in an indeterministic worldview, one would have to reconcile  the apparent need for a discretized time --due to a series of genuine acts of creation when potentiality becomes actuality-- with the continuous geometry of space-time entailed by relativity (see \cite{bisio} for a work in this direction). Note, however, that we do not argue for a fundamental discretization of space-time, we agree that continuity is an irreplaceable feature in relativity theory, but we maintain that the continuum is retrieved similarly to Brouwer's ``viscous continuum'' in intuitionistic mathematics, wherein numbers that constitute the continuum are processes that get more and more determinate as time passes \cite{gisin2020}. Can this proposed approach in terms of finite information help go beyond the standard relativistic block-universe picture, in which time is a mere illusion, and support instead Reichenbach's view according to which ``the parallelism [between space and time] does not exist objectively and that in natural science time is more fundamental than space''? \cite{reichenbach, smolin}.


\acknowledgements
We would like to thank Giacomo Mauro D'Ariano, \v{C}aslav Brukner, Guilherme Franzmann, Luca Apadula and Gemma De las Cuevas for useful discussions and for pointing out relevant literature. F.D.S. acknowledges the financial support through a DOC Fellowship of the Austrian Academy of Sciences (\"OAW). N.G. acknowledges support from the NCCR SwissMap.
\begin{small}

\end{small}

\end{document}